\documentclass[letterpaper]{article}
\usepackage{aaai}
\usepackage{times}
\usepackage{helvet}
\usepackage{courier}
\usepackage{textcomp}
\usepackage{amsmath,amssymb}
\usepackage[pdftex]{graphicx}
\newcommand{\modelname}{PCLF}
\newcommand{\postzero}{P(C_u^{(k)},C_{vcom}^{(t)}\big|u_i,v_i,r_i)}
\newcommand{\postone}{P(C_u^{(k)},C_{vspe1}^{(l_1)}\big|u_i^{(1)},v_i^{(1)},r_i^{(1)})}
\newcommand{\posttwo}{P(C_u^{(k)},C_{vspe2}^{(l_2)}\big|u_i^{(2)},v_i^{(2)},r_i^{(2)})}
\newcommand{\Pzero}{P_0(k,t|j)}
\newcommand{\Pone}{P_1(k,l_1|j)}
\newcommand{\Ptwo}{P_2(k,l_2|j)}
\newcommand{\Cu}[1]{P(C_u^{#1})}
\newcommand{\Cvcom}[1]{P(C_{vcom}^{#1})}
\newcommand{\Cvspeone}[1]{P(C_{vspe1}^{#1})}
\newcommand{\Cvspetwo}[1]{P(C_{vspe2}^{#1})}
\newcommand{\uConCu}[3]{P(u_{#2}^{#1}\big|C_u^{#3})}
\newcommand{\vConCvcom}[2]{P(v_{#1}\big|C_{vcom}^{#2})}
\newcommand{\vConCvspeone}[2]{P(v_{#1}^{(1)}\big|C_{vspe1}^{#2})}
\newcommand{\vConCvspetwo}[2]{P(v_{#1}^{(2)}\big|C_{vspe2}^{#2})}
\newcommand{\rConCuCvcom}[3]{P(r_{#1}\big|C_u^{#2},C_{vcom}^{#3})}
\newcommand{\rConCuCvspeone}[4]{P(r_{#2}^{#1}\big|C_u^{#3},C_{vspe1}^{#4})}
\newcommand{\rConCuCvspetwo}[4]{P(r_{#2}^{#1}\big|C_u^{#3},C_{vspe2}^{#4})}
\newcommand{\rConCuCvspe}[2]{P(r\big|C_u^{#1},C_{vspez}^{#2})}
\newcommand{\CuConu}[1]{P(C_u^{#1}\big|u)}
\newcommand{\CvcomConv}[1]{P(C_{vcom}^{#1}\big|v)}
\newcommand{\CvspeConv}[2]{P(C_{vspe#2}^{#1}\big|v)}
\newcommand{\sumrcom}[2]{\sum_r r\rConCuCvcom{}{#1}{#2}}
\newcommand{\sumrspe}[2]{\sum_r r\rConCuCvspe{#1}{#2}}
\newcommand{\CuEqu}{\Cu{(k)} = \frac{\sum_{t}\sum_{j}\Pzero+\sum_{z}\sum_{l}\sum_{j}P_z(k,l|j)}
{2\times\sum_{z}S_z}}
\newcommand{\CvcomEqu}{\Cvcom{(t)}=\frac{\sum_{k}\sum_{j}\Pzero}{\sum_{z}S_z}}
\newcommand{\CvspeoneEqu}{\Cvspeone{(l_1)}=\frac{\sum_{k}\sum_{j}\Pone}{S_1}}
\newcommand{\CvspetwoEqu}{\Cvspetwo{(l_2)}=\frac{\sum_{k}\sum_{j}\Ptwo}{S_2}}

\newcommand{\vConCvcomEqu}{\vConCvcom{}{(t)}=\frac{\sum_{k}\sum_{j:v_j=v}\Pzero}
{\Cvcom{(t)}\sum_{z}S_z}}
\newcommand{\vConCvspeoneEqu}{\vConCvspeone{}{(l_1)}=\frac{\sum_{k}\sum_{j:v_j^{(1)}=v^{(1)}}\Pone}
{\Cvspeone{(l_1)}\times S_1}}
\newcommand{\vConCvspetwoEqu}{\vConCvspetwo{}{(l_2)}=\frac{\sum_{k}\sum_{j:v_j^{(2)}=v^{(2)}}\Ptwo}
{\Cvspetwo{(l_2)}\times S_2}}
\newcommand{\rConCuCvcomEqu}{\rConCuCvcom{}{(k)}{(t)}=\frac{\sum_{j:r_j=r}\Pzero}{\sum_{j}\Pzero}}
\newcommand{\rConCuCvspeoneEqu}{\rConCuCvspeone{}{}{(k)}{(l_1)}=
\frac{\sum_{j:r_j^{(1)}=r}\Pone}{\sum_{j}\Pone}}
\newcommand{\rConCuCvspetwoEqu}{\rConCuCvspetwo{}{}{(k)}{(l_2)}=
\frac{\sum_{j:r_j^{(2)}=r}\Ptwo}{\sum_{j}\Ptwo}}
\newcommand{\uWithCu}[3]{\Cu{#3}\uConCu{#1}{#2}{#3}}
\newcommand{\vWithCvcom}[2]{\Cvcom{#2}\vConCvcom{#1}{#2}}
\newcommand{\vWithCvspeone}[2]{\Cvspeone{#2}\vConCvspeone{#1}{#2}}
\newcommand{\vWithCvspetwo}[2]{\Cvspetwo{#2}\vConCvspetwo{#1}{#2}}

\newcommand{\postzeroequV}{\frac{P(u_i,C_u^{(k)})P(v_i,C_{vcom}^{(t)})\rConCuCvcom{i}{(k)}{(t)}}
{\sum_{p,q}P(u_i,C_u^{(p)})P(v_i,C_{vcom}^{(q)})\rConCuCvcom{i}{(p)}{(q)}}}
\newcommand{\postoneequV}{\frac{P(u_i^{(1)},C_u^{(k)})P(v_i^{(1)},C_{vspe1}^{(l_1)})\rConCuCvspeone{(1)}{i}{(k)}{(l_1)}}
{\sum_{p,q}P(u_i^{(1)},C_u^{(p)})P(v_i^{(1)},C_{vspe1}^{(q)})\rConCuCvspeone{(1)}{i}{(p)}{(q)}}}
\newcommand{\posttwoequV}{\frac{P(u_i^{(2)},C_u^{(k)})P(v_i^{(2)},C_{vspe2}^{(l_2)})\rConCuCvspeone{(2)}{i}{(k)}{(l_2)}}
{\sum_{p,q}P(u_i^{(2)},C_u^{(p)})P(v_i^{(2)},C_{vspe2}^{(q)})\rConCuCvspeone{(2)}{i}{(p)}{(q)}}}
\begin{document}
\title{Improving Cross-domain Recommendation through\\ Probabilistic Cluster-level Latent Factor Model\\ --- Extended Version}
\author{Siting Ren\and Sheng Gao\\Beijing University of Posts and Telecommunications\\No.10 Xi Tu Cheng Road, Beijing, 100876, China\\rensiting@bupt.edu.cn, gaosheng@bupt.edu.cn\\}
\maketitle
\begin{abstract}
\begin{quote}
Cross-domain recommendation has been proposed to transfer user behavior pattern by pooling together the rating data from multiple domains to alleviate the sparsity problem appearing in single rating domains. However, previous models only assume that multiple domains share a latent common rating pattern based on the user-item co-clustering. To capture diversities among different domains, we propose a novel Probabilistic Cluster-level Latent Factor (PCLF) model to improve the cross-domain recommendation performance. Experiments on several real world datasets demonstrate that our proposed model outperforms the state-of-the-art methods for the cross-domain recommendation task.
\end{quote}
\end{abstract}
\section{Introduction}
Traditional recommender systems based on collaborative filtering (CF) aim to provide recommendations for users on a set of items belonging to only a single domain (e.g., music or
movie) based on the historical user-item preference records. However, CF recommender systems often suffer from the sparsity problem, because in many cases, users rate only a limited number of items, even the item space is often very large. And the sparse rating matrix results in low-quality predictions. With the increasing of user-generated content, there exists a considerable number of publicly available user-item ratings from different domains,
thus, instead of treating items from each single domain independently, knowledge acquired in a single domain could be transferred and shared in other related domains, which has been referred to as \emph{Cross-Domain Recommendation}. \cite{gao2013cross}

Cross-domain recommendation models have shown that knowledge transfer and sharing among the related domains can be beneficial to alleviate the data sparsity problem in single-domain recommendations.
CBT \cite{li2009can} is an early transfer learning algorithm, which studies knowledge transferability between two distinct data. A cluster-level rating pattern (a.k.a., codebook) is constructed from the auxiliary data via some user-item co-clustering algorithm. Then the codebook is transferred to the target data via codebook expansion. A later extension called RMGM \cite{li2009transfer} combines codebook construction and codebook expansion in CBT into one single step with soft membership indicator matrices. Considering the existence of more than one auxiliary data, TALMUD \cite{moreno2012talmud} extends the codebook in CBT to multiple codebooks with different relatedness weight. These models all assume that multiple domains share the common latent rating pattern.

However, related domains do not necessarily share such a common rating pattern. The diversity among the related domains might outweigh the advantages of the common rating pattern, which may result in performance degradations. That is, the existing models cannot consider the domain-specific knowledge about the rating patterns to improve the mutual strengths in cross-domain recommendation.

To learn the shared knowledge and not-shared effect of each domain simultaneously, we propose a novel Probabilistic Cluster-level Latent Factor (PCLF) model to enhance the cross-domain recommendation, which can learn the common rating pattern shared across domains with
the flexibility of controlling the optimal level of sharing, as well as capture the domain-specific rating patterns of users and clustering of items in each domain. Meanwhile, in order to alleviate the sparsity problem in rating datasets, we also construct the priors on users and items by incorporating the user-specific clusters and item-specific clusters. Experiments on several real world datasets show that our proposed model outperforms the state-of-the-art methods for the cross-domain recommendation task. With experiments we offer evidence that our model can do better on alleviating the sparsity problem.

\section{Problem Setting}
Suppose that we are given $Z$ rating matrices from related domains for personalized item recommendation. In the $z$-th domain rating matrix there are a set of users
$U_z = \{u_1^{(z)},\ldots,u_{M_z}^{(z)}\}$
to rate a set of items $V_z = \{v_1^{(z)},\ldots,v_{N_z}^{(z)}\}$, where $M_z$ and $N_z$ represent the numbers of rows (users) and columns (items) respectively. Here the set of users and items across multiple domains may overlap or be isolated with each other. In this work
we consider the more difficult case that neither the users or the items in the multiple rating matrices are overlapping.
The rating data in the $z$-th rating matrix is a set of triplets
$D_z = \{ (u_{1}^{(z)},v_{1}^{(z)},r_{1}^{(z)}),\ldots,(u_{S_z}^{(z)},v_{S_z}^{(z)},r_{S_z}^{(z)})\}$
,where $S_z$ is the number of available ratings in the $z$-th rating matrix.The ratings in {$D_1$,\ldots,$D_Z$} should be in the same rating scales R (e.g.,1-5).

In our cross-domain collaborative filtering setting, we consider how to predict the missing ratings in all domains of interest by transferring correlated knowledge across domains.

\section{Our Proposed Model}
\subsection{Model Specification}
Our Proposed Model is motivated by the following observations. In real-world scenarios, for items, we see the domain-specific clusters and the common clusters exist simultaneously. For example, movies and music can be both classified by regions, but a movie category (e.g., science fiction) may not be able to describe music. Also regions (the common clusters) and the movie category (the domain-specific clusters) affect the rating results of movies in a certain proportion. So items among multiple domains may not always be able to be grouped into high quality clusters, and including a domain-specific rating pattern (user clusters rating on domain-specific item clusters) may be more accurate than sharing the common knowledge only.

\subsection{Cluster-Level Latent Factor}
We assume that the hidden cluster-level structures across domains can be extracted to learn the rating-pattern of user groups on the item clusters for knowledge transfer and sharing, and to clearly demonstrate the co-clusters of users and items.

\subsubsection{User Clusters}
In real world, users may have multiple personalities, so a user can simultaneously
belong to multiple user clusters. Suppose there are $K$ user clusters among all $Z$ domains, $\{C_u^{(1)},C_u^{(2)},\ldots,C_u^{(K)}\}$. The probability of a user $u$ belonging to an exact user cluster $k$ can be described as $\CuConu{(k)}$. Thus, we can define user cluster membership vector for a user $u$ as
\begin{equation}\label{Pu}
  \mathbf{p_u}=\big[\CuConu{(1)},\CuConu{(2)},\ldots,\CuConu{(K)}\big]
\end{equation}
s.t. $\mathbf{p_u 1} = 1$

\subsubsection{Item Clusters}
According to the example in the previous section, we define two types of item clusters: \\
\begin{itemize}
\item[--]\textbf{common item clusters}: which may represent the mutual features of items in all Z domains. Suppose there are $T$ common item clusters, $\{C_{vcom}^{(1)},C_{vcom}^{(2)},\ldots,C_{vcom}^{(T)}\}$. Similar to user clusters, we can define common item cluster membership vector for an item $v$ as
\begin{equation}\label{Pvcom}
  \mathbf{p_{vcom}}=\big[\CvcomConv{(1)},\ldots,\CvcomConv{(T)}\big]
\end{equation}
s.t. $\mathbf{p_{vcom} 1} = 1$\\
\item[--]\textbf{domain-specific item clusters}: which may describe the properties of items in each domain. Suppose there are $L_z$ domain-specific item clusters for the $z$-th domain, $\{C_{vspez}^{(1)},C_{vspez}^{(2)},\ldots,C_{vspez}^{(L_z)}\}$. Also, we define domain-specific item cluster membership vector for an item $v$ in the $z$-th domain as
\begin{equation}\label{Pvspe}
  \mathbf{p_{vspez}}=\big[\CvspeConv{(1)}{z},\ldots,\CvspeConv{(L_z)}{z}\big]
\end{equation}
s.t. $\mathbf{p_{vspez} 1} = 1$
\end{itemize}

\subsubsection{Cluster-Level Rating Matrix}
Furthermore, a user-item co-cluster can also have multiple ratings with different probabilities. For user cluster $i$ rating common item cluster $j$, that is $\rConCuCvcom{}{(i)}{(j)}$. And for user cluster $i$ rating domain-specific item cluster $j$ in the $z$-th domain, that is $\rConCuCvspe{(i)}{(j)}$.

Then we can construct cluster-level rating matrix $S$.\\
Common cluster-level rating matrix $S_{com}\in \mathbb{R}^{K\times T}$ defined as
\begin{multline*}
  S_{com}=\\
  \begin{bmatrix}
    \sumrcom{(1)}{(1)}&\cdots&\sumrcom{(1)}{(T)} \\
    \sumrcom{(2)}{(1)}&\cdots&\sumrcom{(2)}{(T)} \\
    \vdots  & \ddots & \vdots \\
    \sumrcom{(K)}{(1)}&\cdots&\sumrcom{(K)}{(T)}
  \end{bmatrix}
\end{multline*}
Each element $S_{com}(i,j)$ denotes the expectation of rating given by user cluster $i$ to common item cluster $j$.

Similarly, we obtain the definition of domain-specific cluster-level rating matrix $S_{spez}\in \mathbb{R}^{K\times L_z}$
\begin{multline*}
  S_{spez}=\\
  \begin{bmatrix}
    \sumrspe{(1)}{(1)} & \cdots & \sumrspe{(1)}{(L_z)} \\
    \sumrspe{(2)}{(1)} & \cdots & \sumrspe{(2)}{(L_z)} \\
    \vdots  & \ddots & \vdots \\
    \sumrspe{(K)}{(1)}  & \cdots & \sumrspe{(K)}{(L_z)}
  \end{bmatrix}
\end{multline*}
Here, each element $S_{spez}(i,j)$ denotes the expectation of rating given by user cluster $i$ to domain-specific item cluster $j$ in the $z$-th domain.

\subsubsection{Probabilistic Cluster-level Latent Factor Model}
Based on the assumption that random variables $u$ and $v$ are independent \cite{li2009transfer}, we can define the rating function $f_R(u,v)$ for a user $u$ on an item $v$, which can be defined by the two combined rating function: the cross-domain rating function $f_{R_c}(u,v)$ and the domain-specific rating function $f_{R_s}^{(z)}(u,v)$. For that, the cross-domain rating function $f_{R_c}(u,v)$ in terms of the two latent cluster variables $C_u^{(k)}$ and $C_{vcom}^{(t)}$ can be defined as follows:
\begin{align}\label{Frcone}
  f_{R_c}&(u,v) =  \mathbf{p_u}S_{com} \mathbf{p_{vcom}^T}  \\
    &=\sum_r r\sum_{k,t}\rConCuCvcom{}{(k)}{(t)}\CuConu{(k)}\CvcomConv{(t)}\nonumber\\
    &=\sum_r r\sum_{k,t}\rConCuCvcom{}{(k)}{(t)}P(C_u^{(k)},C_{vcom}^{(t)}\big|u,v)\nonumber\\
    &=\sum_r rP(r\big|u,v) \label{Frctwo}
\end{align}
Then the domain-specific rating function $f_{R_s}^{(z)}(u,v)$ in terms of the two latent cluster variables $C_u^{(k)}$ and $C_{vspez}^{(l_z)}$ in the $z$-th domain can be written as follows:
\begin{align}\label{Frsone}
  f_{R_s}^{(z)}&(u,v) =  \mathbf{p_u}S_{spez} \mathbf{p_{vspez}^T}  \\
    &=\sum_r r\sum_{k,l_z}\rConCuCvspe{(k)}{(l_z)}\CuConu{(k)}\CvspeConv{(l_z)}\nonumber\\
    &=\sum_r r\sum_{k,l_z}\rConCuCvspe{(k)}{(l_z)}P(C_u^{(k)},C_{vspez}^{(l_z)}\big|u,v)\nonumber\\
    &=\sum_r rP(r\big|u,v) \label{Frstwo}
\end{align}
Equations above implies that $f_{R_c}(u,v)$ gives ratings from the perspective of cross-domain recommendation. While, $f_{R_s}^{(z)}(u,v)$ gives ratings from the perspective of single-domain recommendation. Based on the idea that our model combines cross-domain recommendation with single-domain recommendation, we introduce a set of variables to balance the effect between them. For the $z$-th domain, set $W_1^{(z)}$ to be the weight of cross-domain recommendation, and $W_2^{(z)}$ to be the weight of single-domain recommendation. s.t. $W_1^{(z)}+W_2^{(z)}=1$.

Thus, we can define the rating function $f_R(u,v)$ for the $z$-th domain as
\begin{align}\label{Fr}
  f_R&(u,v)= W_1^{(z)}f_{R_c}(u,v)+W_2^{(z)}f_{R_s}^{(z)}(u,v)\\
   &=W_1^{(z)}(\mathbf{p_u}S_{com} \mathbf{p_{vcom}^T})+W_2^{(z)}(\mathbf{p_u}S_{spez} \mathbf{p_{vspez}^T})\nonumber
\end{align}
\begin{figure*}
  \centering
  \includegraphics[width=6.5in]{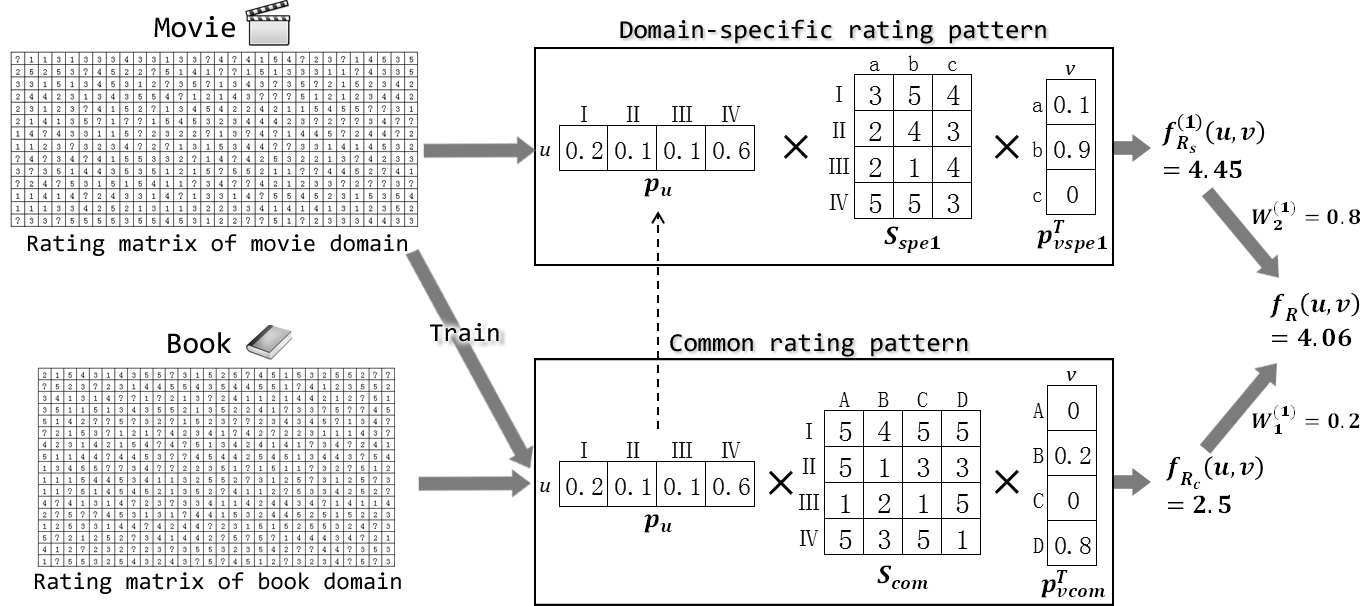}\\
  \caption{Illustration of our proposed \modelname{} model in the context of two related domains. The rating prediction result $f_R(u,v)$ of a user $u$ on an item $v$ in movie domain is the combination of cross-domain recommendation result $f_{R_c}(u,v)$ and single-domain recommendation result $f_{R_s}^{(1)}(u,v)$ with specific weight $W_1^{(1)}$ and $W_2^{(1)}$.}
  \label{illustration}
\end{figure*}
The illustration of our proposed \modelname{} model can be found in Figure \ref{illustration}.

\section{Model Learning}
In this section, we introduce how to train our model on the pooled rating data $\bigcup_z D_z$. The Expectation and Maximization (EM) \cite{dempster1977maximum} algorithm is a well-known optimization
algorithm, which alternates between two steps: the expectation step and the maximization step. Here we adopt EM algorithm for model training.

For ease of understanding and without loss of generality, we set $Z$=2, i.e., two domains for recommendation. We need to learn eleven sets of model parameters, i.e., $\Cu{(k)}$, $\Cvcom{(t)}$, $\Cvspeone{(l_1)}$, $\Cvspetwo{(l_2)}$, $\uConCu{}{}{(k)}$, $\vConCvcom{}{(t)}$, $\vConCvspeone{}{(l_1)}$, $\vConCvspetwo{}{(l_2)}$,  $\rConCuCvcom{}{(k)}{(t)}$, $\rConCuCvspeone{}{}{(k)}{(l_1)}$, $\rConCuCvspetwo{}{}{(k)}{(l_2)}$. For $k=1,\ldots,K$; $l_1=1,\ldots,L_1$; $l_2=1,\ldots,L_2$; $t=1,\ldots,T$; $u\in \bigcup_z U_z$; $v\in \bigcup_z V_z$ and $r\in R$.

\subsubsection{Expectation step}

The joint posterior probabilities are computed as:
\begin{multline}\label{postzero}
  \postzero = \\
  \postzeroequV
\end{multline}
\begin{multline}\label{postone}
  \postone = \\
  \postoneequV
\end{multline}
\begin{multline}\label{posttwo}
  \posttwo= \\
  \posttwoequV
\end{multline}
Where Equation \eqref{postzero} is computed using the pooled rating data $\bigcup_z D_z$.
Equation \eqref{postone} is computed using the rating data in the first rating matrix $D_1$, and Equation \eqref{posttwo} is computed using the rating data in the second rating matrix $D_2$.\\
And $P(u_i^{(z)},C_u^{(k)})=\uWithCu{(z)}{i}{(k)}$ for $z=1,2$\\
$P(v_i,C_{vcom}^{(t)})=\vWithCvcom{i}{(t)}$\\
$P(v_i^{(1)},C_{vspe1}^{(l_1)})=\vWithCvspeone{i}{(l_1)}$\\
$P(v_i^{(2)},C_{vspe2}^{(l_2)})=\vWithCvspetwo{i}{(l_2)}$\\

\subsubsection{Maximization step}

For simplicity,
let $\Pzero$, $\Pone$, $\Ptwo$ as shorthands for $\postzero$, $\postone$ and $\posttwo$, respectively.
Then the model parameters are updated as:
\begin{equation}\label{Cu}
  \CuEqu
\end{equation}
\begin{equation}\label{Cvcom}
  \CvcomEqu
\end{equation}
\begin{equation}\label{Cvspeone}
  \CvspeoneEqu
\end{equation}
\begin{equation}\label{Cvspetwo}
  \CvspetwoEqu
\end{equation}
\begin{equation}\label{uConCu}
  \uConCu{}{}{(k)}=
\end{equation}
\begin{small}
  $$[\sum_{t}\sum_{j:u_j=u}\Pzero+\sum_{l_1}\sum_{j:u_j^{(1)}=u}\Pone$$\\
  $$+\sum_{l_2}\sum_{j:u_j^{(2)}=u}\Ptwo]\Big/ [\Cu{(k)}(2\times\sum_{z}S_z)]$$
\end{small}

\begin{equation}\label{vConCvcom}
  \vConCvcomEqu
\end{equation}
\begin{equation}\label{vConCvspeone}
  \vConCvspeoneEqu
\end{equation}
\begin{equation}\label{vConCvspetwo}
  \vConCvspetwoEqu
\end{equation}
\begin{equation}\label{rConCuCvcom}
  \rConCuCvcomEqu
\end{equation}
\begin{equation}\label{rConCuCvspeone}
  \rConCuCvspeoneEqu
\end{equation}
\begin{equation}\label{rConCuCvspetwo}
  \rConCuCvspetwoEqu
\end{equation}

To avoid the local maximum problems, we use a general form of the EM algorithm
named annealed EM algorithm (AEM) \cite{hofmann1998statistical}, which is an EM algorithm with regularization. We adopt the same method used in FMM \cite{si2003flexible} for applying AEM to training procedure.
\subsubsection{Model Inference}
After training the model, we get those sets of model parameters. Hence, the following cluster-level rating matrix can be obtained:
\begin{equation}\label{Scom}
  S_{com}(i,j)=\sum_r r\rConCuCvcom{}{(i)}{(j)}
\end{equation}
for $i=1,\ldots,K; j=1,\ldots,T$
\begin{equation}\label{Sspeone}
  S_{spe1}(i,j)=\sum_r r\rConCuCvspeone{}{}{(i)}{(j)}
\end{equation}
for $i=1,\ldots,K; j=1,\ldots,L_1$
\begin{equation}\label{Sspetwo}
  S_{spe2}(i,j)=\sum_r r\rConCuCvspeone{}{}{(i)}{(j)}
\end{equation}
for $i=1,\ldots,K; j=1,\ldots,L_2$

Also, the following parameters can be computed using the learned parameters based on the Bayes rule:
\begin{equation}
  \CuConu{(k)}=\frac{\uConCu{}{}{(k)}\Cu{(k)}}{\sum_k \uConCu{}{}{(k)}\Cu{(k)}}
\end{equation}
\begin{equation}
  \CvcomConv{(t)}=\frac{\vConCvcom{}{(t)}\Cvcom{(t)}}{\sum_t \vConCvcom{}{(t)}\Cvcom{(t)}}
\end{equation}
\begin{equation}
  P(C_{vspe1}^{(l_1)}\big|v^{(1)})=
  \frac{\vConCvspeone{}{(l_1)}\Cvspeone{(l_1)}}
  {\sum_{l_1}\vConCvspeone{}{(l_1)}\Cvspeone{(l_1)} }
\end{equation}
\begin{equation}
  P(C_{vspe2}^{(l_2)}\big|v^{(2)})=
  \frac{\vConCvspetwo{}{(l_2)}\Cvspetwo{(l_2)}}
  {\sum_{l_2}\vConCvspetwo{}{(l_2)}\Cvspetwo{(l_2)} }
\end{equation}

According to Equation \eqref{Fr}, missing values in the first rating matrix can be generated by
\begin{equation*}
  f_R(u,v) =W_1^{(1)}(\mathbf{p_u}S_{com} \mathbf{p_{vcom}^T})+W_2^{(1)}(\mathbf{p_u}S_{spe1} \mathbf{p_{vspe1}^T})
\end{equation*}
And missing values in the second rating matrix can be generated by
\begin{equation*}
  f_R(u,v) =W_1^{(2)}(\mathbf{p_u}S_{com} \mathbf{p_{vcom}^T})+W_2^{(2)}(\mathbf{p_u}S_{spe2} \mathbf{p_{vspe2}^T})
\end{equation*}

Moreover, our model can also predict the ratings on an item in one domain for a user in another domain.

\section{Experiments}
In this section, we examine how our proposed model behaves on real-world rating datasets and compare it with several state-of-the-art single-domain recommendation models and cross-domain recommendation models:
\begin{itemize}
\item[--]NMF (Nonnegative Matrix Factorization) \cite{seung2001algorithms}: a single-domain
model which employs nonnegative matrix factorization method to learn the
latent factors in each domain and provide the prediction performance separately.
\item[--]FMM (Flexible Mixture Model) \cite{si2003flexible}: a single-domain model which uses probabilistic mixture model to learn latent cluster structure in each single domain and then provide the single domain performance separately.
\item[--]RMGM (Rating-Matrix Generative Model) \cite{li2009transfer}: a cross-domain model which can only transfer and share the common rating pattern by the cluster-level rating matrix across multiple domains.
\item[--]\modelname{} model: our proposed model.
\end{itemize}

\subsection{Datasets}
For the experiments we have used the following benchmark real-world datasets for performance
evaluation:
\begin{itemize}
\item[--]\textit{MovieLens} dataset\footnote{http://www.grouplens.org/node/73} : contains more than 100,000 movie ratings with the scales from 1 to 5 provided by 943 users on 1,682 movies. We randomly choose 500 users with more than 16 ratings and 1000 movies for experiments.
\item[--]\textit{EachMovie} dataset\footnote{http://www.cs.cmu.edu/$\sim$lebanon/IR-lab.htm} : contains 2.8 million movie ratings with the scales from 1 to 6 provided by 72,916 users on 1,628 movies. We also randomly choose 500 users with more than 20 ratings and 1000 movies for
    experiments. Here, we map 6 to 5 to normalize the rating scales from 1 to 5.
\item[--]\textit{Book-Crossing} dataset\footnote{http://www.informatik.uni-freiburg.de/$\sim$cziegler/BX/} : contains more than 1.1 million ratings with the scales from 0 to 9 provided by 278,858 users on 271,379 books. We still randomly select 500 users and 1000 books with more than 16 ratings for each item in the experiments. We also normalize the rating scales from 1 to 5 for fair comparison.

\end{itemize}
\subsection{Evaluation Protocol}
We examine the compared models on different datasets under different configurations. The first 300 users in this three datasets are used for training, respectively, and the last 200 users for testing. For each test user, we consider to keep different sizes of the observed ratings as
the initialization of each user in the experiments, i.e., 5, 10 or 15 ratings of each test
user are given to avoid cold-start problem and the remaining ratings are used for evaluation. For example, each test user keeps 10 observed ratings in the \textit{MovieLens} training set is expressed as ML-Given10.

We choose \textit{Book-Crossing} vs \textit{EachMovie} and \textit{Book-Crossing} vs \textit{MovieLens} and \textit{MovieLens} vs \textit{EachMovie} as three kinds of related domains to discover the relatedness among different domains. In the experiments we conduct the methods by repeating the process 10 times and report the average results.

To check the performances of different methods, we use MAE (Mean Absolute Error) as the evaluation
metric. MAE is computed as $MAE=\sum_{i\in O}|r_i-r_i^*|/|O|$, where $|O|$ denotes the number of test ratings, $r_i$ is the true value and $r_i^*$ is the predicted rating. The smaller the value of MAE is, the better the model performs.

\subsection{Experiment Results}
Since we use EM algorithm for model training, and the performances of EM algorithm is sensitive to initialization. So, the influence of a good initialization to our model performances is significant. After testing different initialization methods, we finally use random values for initializing $\postzero$, $\postone$ and $\posttwo$. Note that they should also be respectively normalized. Then we can obtain the eleven sets of initialized parameters according to Equation \eqref{Cu}-\eqref{rConCuCvspetwo}.

The parameters of different models have been manually tuned and we report here the best results obtained based on the optimal combination of many parameter settings. For our model, we observed that the performance is rather stable when $K$, $L_1$, $L_2$ and $T$ are in the range of [10,50].

\begin{figure}
  \centering
  \includegraphics[width=3.3in]{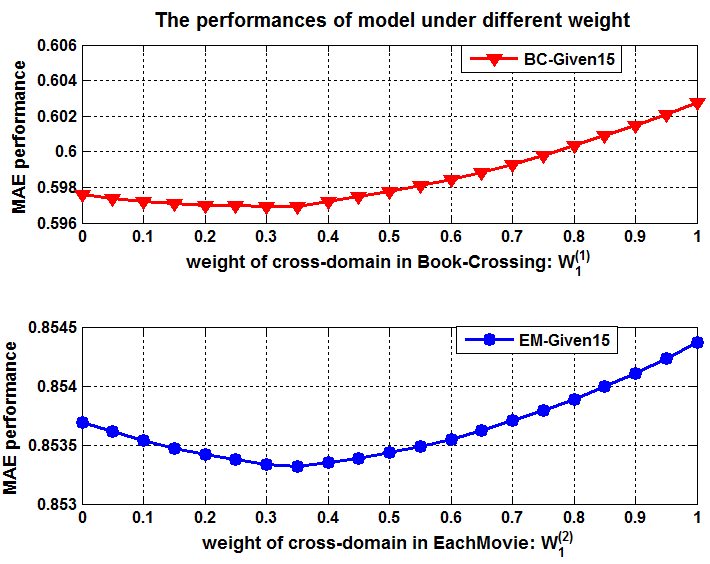}\\
  \caption{The performances of model under different weight}
  \label{weight}
\end{figure}

Table \ref{BCEM} shows the MAE performances of the compared models on \textit{Book-Crossing} vs \textit{EachMovie} domains under different configurations, where we set the number of users and item clusters to $K$=20, $L_1$=15, $L_2$=15 and $T$=10 respectively. Figure \ref{weight} provides the performances of our proposed model under different weight configurations, which clearly shows the tradeoff between cross-domain recommendation
and domain-specific recommendation. Thus, in the experiments we choose $W_1^{(1)}=W_1^{(2)}=0.35$, $W_2^{(1)}=W_2^{(2)}=0.65$.

In the experiments, we have 5, 10 and 15 ratings of each test user in the \textit{Book-Crossing} and \textit{EachMovie} datasets that are given for training while the remaining ratings are used for test, and the compared models are evaluated on the different combined settings as BC-Given5 vs EM-Given5, BC-Given10 vs EM-Given10 and BC-Given15 vs EM-Given15. And we conduct the same combined settings on \textit{Book-Crossing} vs \textit{MovieLens} and \textit{MovieLens} vs \textit{EachMovie}. The experiment results are reported in Table \ref{BCEM},\ref{BCML},\ref{MLEM}. Best results are in bold.

From the results we can see that the best performing method among all the models is our proposed model. We also observed that the cross-domain based models clearly outperforms the single domain based models, which shows that the latent cross-domain common rating pattern can indeed aggregate more useful information than the single-domain methods do individually. Moreover, our proposed \modelname{} model provides even better results than the state-of-the-art cross-domain recommendation model RMGM, which indicates the benefits of combining the cross-domain information with the domain-specific knowledge to enhance the cross-domain recommendation accuracy.

\begin{table}
  \centering
  \begin{tabular}{cc|ccc}
    \hline
    \hline
    Dataset & Model & Given 5 & Given 10 & Given 15 \\
    \hline
       & NMF & 0.6575 & 0.6375 & 0.6301\\
    BC & FMM & 0.6451 & 0.6196 & 0.6125\\
       & RMGM & 0.6378 & 0.6115 & 0.6092\\
       & \modelname & \textbf{0.6252} & \textbf{0.5994} & \textbf{0.5969}\\
    \hline
       & NMF & 0.9345 & 0.8861 & 0.8799\\
    EM & FMM & 0.9132 & 0.8831 & 0.8771 \\
       & RMGM & 0.9021 & 0.8743 & 0.8637\\
       & \modelname & \textbf{0.8838} & \textbf{0.8677} & \textbf{0.8533}\\
    \hline
    \hline
  \end{tabular}
  \caption{MAE performances of the compared models on \textit{Book-Crossing} vs \textit{EachMovie} related domains under different configurations. The combined settings BC-Given5 vs EM-Given5, BC-Given10 vs EM-Given10 and BC-Given15 vs EM-Given15 are conducted. Best results are in bold.}
  \label{BCEM}
\end{table}

\begin{table}
  \centering
  \begin{tabular}{cc|ccc}
  \hline
  \hline

  Dataset & Model & Given 5 & Given 10 & Given 15 \\
  \hline
     & NMF & 0.6575	& 0.6375 & 0.6301\\
  BC & FMM & 0.6451 & 0.6196 & 0.6125 \\
     & RMGM & 0.6364 & 0.6054 & 0.5994 \\
     & \modelname & \textbf{0.6284} & \textbf{0.5990} & \textbf{0.5950}\\
  \hline
     & NMF & 0.8365 & 0.8016 & 0.7933\\
  ML & FMM & 0.8052 & 0.7838 & 0.7721\\
     & RMGM & 0.8039 & 0.7769 & 0.7692\\
     & \modelname & \textbf{0.7922} & \textbf{0.7729} & \textbf{0.7653}\\
  \hline
  \hline
  \end{tabular}
  \caption{MAE performances of the compared models on \textit{Book-Crossing} vs \textit{MovieLens} related domains under different configurations. The combined settings BC-Given5 vs ML-Given5, BC-Given10 vs ML-Given10 and BC-Given15 vs ML-Given15 are conducted.}
  \label{BCML}
\end{table}

\begin{table}
  \centering
  \begin{tabular}{cc|ccc}
    \hline
    \hline
    Dataset & Model & Given 5 & Given 10 & Given 15 \\
    \hline
       & NMF & 0.8365 & 0.8016 & 0.7933\\
    ML & FMM & 0.8052 & 0.7838 & 0.7721\\
       & RMGM & 0.7985 & 0.7736 & 0.7677\\
       & \modelname & \textbf{0.7878} & \textbf{0.7724} & \textbf{0.7639}\\
    \hline
       & NMF & 0.9345 & 0.8861 & 0.8799\\
    EM & FMM & 0.9132 & 0.8831 & 0.8771\\
       & RMGM & 0.9017 & 0.8733 & 0.8601\\
       & \modelname & \textbf{0.8820} & \textbf{0.8658} & \textbf{0.8498}\\
    \hline
    \hline
  \end{tabular}
  \caption{MAE performances of the compared models on \textit{MovieLens} vs \textit{EachMovie} related domains under different configurations. The combined settings ML-Given5 vs EM-Given5, ML-Given10 vs EM-Given10 and ML-Given15 vs EM-Given15 are conducted.}
  \label{MLEM}
\end{table}

Furthermore, from Table \ref{BCML} and Table \ref{MLEM} we can also discover that the performances for the item recommendation in the \textit{MovieLens} dataset are not identical even in terms of the same users and items when combined with different related domains in the experiments. The results show that different domains may have various levels of shared information, which are underlying across domains.

\section{Conclusion}
In this work, we proposed a novel cross-domain collaborative filtering method, named probabilistic cluster-level latent factor (\modelname) model. The \modelname{} model has taken into account both the useful knowledge across multiple related domains and the structures in each domain. On the one hand, the \modelname{} model is able to learn shared common rating pattern across multiple rating matrices to alleviate the sparsity problems in individual domain. On the other hand, our model can also draw the discriminative information from each domain to construct a set of latent space to represent the domain-specific rating patterns of user groups on the item clusters from each domain, which is propitious to the improvement of recommendation accuracy. The experimental results show that our proposed \modelname{} model indeed can benefit from the combination of two types of cluster-level rating patterns (i.e., common \& domain-specific) and outperforms the state-of-the-art methods for cross-domain recommendation task.

There are still several extensions to improve our work. At present, we only conduct experiments on datasets with explicit rating (e.g., from 1 to 5). We plan to explore the ability to handle the implicit preferences of users (e.g., visit, click or comment) of our model. Also, most recommendation systems are expected to handle enormous amounts of data (``Big Data") at a reasonable time. So we will evaluate the scalability of our model.

\bibliographystyle{aaai}
\bibliography{Ref}
\nocite{*}

\end{document}